\documentclass[aip,jcp,amsmath,amssymb,reprint]{revtex4-1}
\usepackage{geometry}
\usepackage{amsmath}
\usepackage{amssymb}
\usepackage{graphicx}

\usepackage{xcolor}
\usepackage{enumerate}

\def\br{{\bf r}}
\def\bp{{\bf r^\prime}}

\def\dd{\rm d}

\begin{document}

\title{Stochastic Time-Dependent DFT with Optimally Tuned Range-Separated
Hybrids: Application to Excitonic Effects in Large Phosphorene Sheets}

\author{Vojt\v{e}ch Vl\v{c}ek}
\email{vlcek@ucsb.edu}

\affiliation{Department of Chemistry and Biochemistry, University of California,
Santa Barbara California 93106, U.S.A.}

\author{Roi Baer}
\email{roi.baer@huji.ac.il}

\affiliation{Fritz Haber Center for Molecular Dynamics, Institute of Chemistry,
The Hebrew University of Jerusalem, Jerusalem 91904, Israel}

\author{Daniel Neuhauser}
\email{dxn@ucla.edu}

\affiliation{Department of Chemistry and Biochemistry, University of California,
Los Angeles California 90095, U.S.A.}
\begin{abstract}
We develop a stochastic approach to time-dependent DFT with optimally-tuned
range-separated hybrids containing non-local exchange, for calculating
optical spectra. The attractive electron-hole interaction, which leads
to the formation of excitons, is included through a time-dependent
linear-response technique with a non-local exchange interaction which
is computed very efficiently through a stochastic scheme. The method
is inexpensive and scales quadratically with the number of electrons,
at almost the same (low) cost of time dependent Kohn-Sham (TDKS) with
local functionals. Our results are in excellent agreement with experimental
data and the efficiency of the approach is demonstrated on large finite
phosphorene sheets containing up to 1958 valence electrons. 
\end{abstract}
\maketitle

\section{Introduction}

The understanding of optical excitations in materials is essential
for developing novel optical and electronic devices.\cite{OnidaReiningRubio2002,marques2004time}
However, it is very challenging to calculate optical properties in
extended systems. For small molecules, highly correlated approaches
are used, including configuration interaction or the equation of motion
coupled cluster approach.\cite{bartlett2007coupled,helgaker2014molecular}
Further, the optical response is typically described by the Bethe-Salpeter
equation.\cite{OnidaReiningRubio2002,martin2016interacting} These
high level techniques are predictive but scale steeply with the number
of electrons so they can only be used for relatively small molecules
and unit cells.

An alternative to costly many-body techniques is time-dependent density
functional theory (TDDFT)\cite{Runge1984} that describes excited
state energies, geometries, and other properties of small molecules
with a relatively moderate computational cost. In principle TDDFT
is exact, but in practice approximations have to be introduced. The
most common is the adiabatic time-dependent Kohn-Sham theory (TDKS).
While TDKS has been applied successfully to a wide range of molecular
systems,\cite{OnidaReiningRubio2002,marques2004time} it suffers
from many failures, particularly for extended systems, charge-transfer
excited states,\cite{dreuw2003long} multiple excitations,\cite{maitra2004double} and avoided crossings.\cite{levine2006conical} The most
notable problem of TDKS is the inability to capture low-lying excitonic
states in bulk.\cite{OnidaReiningRubio2002,martin2016interacting}

It has been argued that a TDDFT formulation beyond the Kohn-Sham picture, namely, a TD-GKS (Generalized Kohn-Sham) approach \cite{baer2018time}
which employs a non-local exchange interaction,\cite{Seidl1996,BaerNeuhauser2005,KuemmelKronik2008}
captures the necessary physics to describe excitation in extended
systems \cite{yang2015simple,Refaely-Abramson2015,brawand2016generalization}
and accurately predicts the formation of bound excitons. However,
the inclusion of the non-local exchange in the TDDFT calculations makes them
computationally demanding and out of reach for large nanoscale systems.
Recently a family of stochastic orbital methods has been developed
to describe ground and excited states with the goal of lowering the
computational complexity at the cost of introducing a controllable
statistical error. \cite{BaerNeuhauserRabani2013,Neuhauser2014,Gao2015,Neuhauser2015,Rabani2015,vlcek2017stochastic}
Specifically relevant to the current work is a stochastic ground state
range-separated hybrid DFT method \cite{Neuhauser2015} and a stochastic
method for the Bethe-Salpeter equation (BSE). \cite{Rabani2015}

Our first and main aim in this work is to  overcome, using stochastic methods, the computational bottleneck in TDDFT with non-local exchange. Specifically, we develop a real-time
generalized Kohn-Sham method based on a range-separated hybrid (RSH)
with a long-range exact exchange operator. The approach has similarities
to that proposed for stochastic BSE (where a damped exchange operator
was used), but the starting point is different. Here, the starting
point is long-range-corrected RSH-DFT, a method which is known to
produce excellent charge-transfer states.\cite{Stein2009} Therefore the present long-range
exchange TDDFT starts only from a long-range DFT calculation (which
is also implemented stochastically), so the resulting approach is
self-contained in the DFT/TDDFT framework and does not resort to a
separate computation of individual quasiparticle states; this is in
contrast to a BSE work where the starting point is a prior calculation
of the quasiparticle states which is achievable, e.g., through the
stochastic $GW$ method.\cite{Neuhauser2014,vlcek2017stochastic,vlvcek2018swift} 

The second aim of the paper is then to use the resulting fully ab-initio
stochastic TDDFT method for describing optical excitations in extremely
large systems with thousands of electrons. 

Below, we first (Sec. II) review the basic theory and present our
stochastic implementation. In Sec. III we verify our method by comparing
with experiment for ${\rm PH_{3}}$ and the method is then applied
to study 2D phosphorene sheets of increasing sizes. Conclusions follow
in Sec. IV.

\section{Theory}

It is well-known that absorption can be determined, in linear response,
from the time-evolution of an induced dipole density (see Appendix
I). This time evolution is in principle governed by the time-dependent
Schr\"odinger equation, which is however intractable beyond few-electron
systems. DFT\cite{HohenbergKohn} is practical approach for recasting the many-electron system
as a set of virtual particles interacting via a mean-field exchange-correlation
(xc) potential.

The original formulation of the Kohn-Sham (KS) \cite{KohnSham} DFT
scheme describes the xc interactions by a local potential which is
in principle non-local in time. In practice it is further approximated,
e.g., by (semi)local functionals of the electronic density at given
time (i.e., the functional is adiabatic). As mentioned in the introduction,
this formulation has some notable failures, e.g., missing excitonic
effects. An alternative route, which we pursue here, is to employ
a GKS scheme\cite{Seidl1996,Savin1995,baer2018time} with
non-local long-range asymptotic behavior as required for correct description
of charge transfer and polarizability \cite{ghosez1997long} and for
electron-hole bound states.\cite{BaerNeuhauser2005,yang2015simple,Refaely-Abramson2015,brawand2016generalization}

We first review below the formulation of DFT and TDDFT with long-range
non-local exchange, followed by the details of stochastic implementation.

\subsection{DFT with long-range non-local exchange}

The GKS Hamiltonian reads 
\begin{equation}
H[n,\rho]\equiv h_{0}+v_{H}\left[n\left(\br\right)\right]+v_{C}^{\gamma}\left[n\left(\br\right)\right]+X^{\gamma}\left[\rho\left(\br,\bp\right)\right],\label{Hamiltonian}
\end{equation}
where $h_{0}$ contains the kinetic energy and the electron-nuclear
attraction. The density-density repulsion is given by the Hartree
potential $v_{H}$, and $v_{C}^{\gamma}$ (where $\gamma$ is defined
shortly) is a (semi)local correlation density functional \textendash{}
we use here a local functional form.\cite{PerdewWang} The non-local
exchange interaction $X^{\gamma}$ is a functional of the density-matrix
$\rho\left(\br,\bp\right)$, where the density is of course $n(\br)=\rho(\br,\br).$
(Note that we use different symbols, $n$ and $\rho$, for the density
and density matrix since later we calculate the two separately - one
deterministically and the other stochastically.)

The form of $X^{\gamma}$ derives from a screened Coulomb interaction
governed by a single parameter $\gamma$.\cite{Savin1995,Leininger1997,BaerNeuhauser2005}
Specifically, the Coulomb kernel is partitioned as: 
\begin{equation}
\frac{1}{r}=\frac{{\rm erfc}\left(\gamma r\right)}{r}+\frac{{\rm erf}\left(\gamma r\right)}{r},\label{range-separation-kernel}
\end{equation}
where $\gamma$ is the range separation parameter. The first term
dominates at small distances ($r\to0$) and its contribution to the
exchange is approximated by a local density functional.\cite{PerdewWang,Leininger1997,Livshits2007}
The second term in Eq.~(\ref{range-separation-kernel}) is active
at large distances and gives the non-local Fock exchange ($X_{{\rm nl}}^{\gamma}$).
The matrix element of the exchange vector is a direct product of the
density matrix and the non-local Coulomb interaction (the simple product
form is important later in the stochastic formulation): 
\begin{equation}
X_{{\rm nl}}^{\gamma}(\br,\bp)\equiv-\nu(\br,\bp)\rho\left(\br,\bp\right).\label{Xdeterm}
\end{equation}
Here, the long-range interaction is $\nu(\br,\bp)={\rm erf}\left(\gamma|\br-\bp|\right)/|\br-\bp|$,
and the density matrix is determined from the eigenstates $\rho(\br,\bp)=\sum_{i}f_{i}\phi_{i}(\br)\phi_{i}(\bp)$,
where $i$ is a state and spin index and $f_{i}$ are the occupation
factors, while the density is $n(\br)=\rho(\br,\br)=\sum_{i}f_{i}|\phi_{i}(\br)|^{2}$.
(In the following we do not denote spin explicitly.) Thus, deterministically,
the nonlocal exchange term acts on a general function $\psi$ as 
\begin{align}
 & \left\langle \br\middle|X_{{\rm nl}}^{\gamma}\middle|\psi\right\rangle =\nonumber \\
 & -\sum_{i}f_{i}\phi_{i}(\br)\int\nu(\br,\bp)\phi_{i}\left(\bp\right)\psi\left(\bp\right)\dd\bp.
\end{align}

In the first (DFT) stage, the occupied eigenstates $\phi_{i}(\br)$
of Eq. (\ref{Hamiltonian}) are calculated self-consistently, ensuring
$H\phi_{i}=\varepsilon_{i}\phi_{i},$ where the density and density
matrix are functions of the eigenstates. The value of $\gamma$ is
found by enforcing the IP theorem that requires that the HOMO energy
equals the ionization energy. This optimal tuning leads to good IPs
and fundamental band gaps in finite systems.\cite{Stein2010,Stein2012,Kronik2012}

The method's success stems from the combination of (semi)local functionals
that capture correlation effects well at short distances with the
nonlocal $X_{{\rm nl}}^{\gamma}$ that guarantees for finite systems
the asymptotically correct $1/r$ behavior of the exchange potential
which is crucial for proper inclusion of the attractive electron-hole
interaction.\cite{ghosez1997long,OnidaReiningRubio2002}

\subsection{TDDFT with long-range non-local exchange}

It is well-known (Appendix I) that the absorption spectrum is obtained
from a linear-response propagation of the density. Specifically, for
polarized excitation along a unit vector $\hat{\boldsymbol{{\rm e}}}$,
we apply a small perturbation $\delta\nu(\br,t)=(\br\cdot\hat{\boldsymbol{{\rm e}}})\delta(t)\Delta$
where $\Delta$ is a small constant (typically between $10^{-3}$and
$10^{-5}$ a.u.). Then, the system evolves under the time-dependent
GKS equation (using $\hbar=1$)
\begin{equation}
i \left|\dot{\phi}_{i}\left(t\right)\middle\rangle=\middle[H\left[n\left(t\right),\rho\left(t\right)\right]+\delta\nu(\br,t)\middle]\middle|\phi_{i}\left(t\right)\right\rangle.\label{tprop}
\end{equation}
To simplify the notation we usually do not denote the dependence of
the density and density matrix (and therefore of the time-dependent
Hamiltonian) on the excitation strength $\Delta$.

The Hamiltonian $H\left(t\right)$ is time dependent as it explicitly
depends on the propagated eigenstates $\phi_{i}\left(\br,t\right)$,
the time-dependent charge density $n\left(\br,t\right)=\sum_{i}f_{i}|\phi\left(\bp,t\right)|^{2}$,
and the charge density matrix, $\rho\left(\br,\bp,t\right)=\sum_{i}f_{i}\phi\left(\br,t\right)\phi_{i}^{*}\left(\bp,t\right)$.
The dipole moment along the excitation direction is then calculated
from the density, $\mu(t)=\int(\br\cdot\hat{\boldsymbol{{\rm e}}})n(\br,t)d\br,$
and the absorption spectrum is calculated by Fourier transforming
the dipole moment $\mu(t)$ (Appendix I).

In principle, the exchange-correlation term in the Hamiltonian should
account for memory effects, but since its form is unknown, we resort
to the adiabatic approximation and construct the xc terms directly
from $n\left(\br,t\right)$ and $\rho\left(\br,\bp,t\right)$. Thus,
the difference from Kohn-Sham type adiabatic TDDFT is only in the
exchange kernel.

The application of the non-local exchange as presented in Eq.~(\ref{Xdeterm})
is computationally demanding, due to the integral over the density
matrix. Practical calculations are therefore limited to systems with
a low number of states.

\subsection{Stochastic DFT with non-local exchange}

Next we review our implementation \citep{Neuhauser2015} of the DFT
equations with a stochastic representation of the non-local exchange
operator. This is followed by implementation of stochastic TDDFT in
the next section.

The first step is the DFT ground-state calculation, where we use the
stochastic-exchange approach of Ref.~\onlinecite{Neuhauser2015}.
This grid-based method is done by two key parts. The first is the
representation of the density matrix as an average over stochastic
correlation functions. Specifically, we construct stochastic states
$\left\{ \eta\right\} $, each of which is a linear combination of
all the occupied eigenstates $\left\{ \phi_{i}\right\} $ (cf.~Refs.\onlinecite{Baer2012,Neuhauser2013,BaerNeuhauserRabani2013,Gao2015,Rabani2015})
\begin{equation}
\eta\left(\br\right)=\sum_{i}\sqrt{f_{i}}\phi_{i}\left(\br\right)\langle\phi_{i}|\bar{\eta}\rangle,\label{create_eta-2}
\end{equation}
where $\bar{\eta}$ is a completely random real vector, e.g., $\bar{\eta}(\br)=\pm(dV)^{-1/2},$
and $dV$ is the grid volume-element. It is straightforward to show
that as an operator, the density matrix becomes an average over the
separable terms

\begin{equation}
\rho=\left\{ |\eta\rangle\langle\eta|\right\} _{\bar{\eta}}\label{eq:stoch_densm}
\end{equation}
i.e., $\rho\left(\br,\bp\right)=\left\{ \eta\left(\br\right)\eta\left(\bp\right)\right\} _{\bar{\eta}},$
where $\left\{ \cdots\right\} _{\bar{\eta}}$ denotes a statistical
average over all random states $\bar{\eta.}$ Since the average of
$|\eta\rangle\langle\eta|$ yields the density matrix, we can view
$\eta(\br)$ as a stochastic density amplitude.

In the ground-state DFT stage we supplement the stochastic representation
of the density matrix by a similar stochastic decomposition of the
long-range Coulomb interaction (Eq.~(\ref{eq:stoch_densm})) using stochastic states $\zeta$: 
\begin{equation}
\frac{{\rm erf}\left(\gamma\left|\br-\bp\right|\right)}{\left|\br-\bp\right|}=\left\{ \zeta\left(\br\right)\zeta\left(\bp\right)\right\} _{\bar{\theta}}\label{coul_decomp}
\end{equation}
that are evaluated as 
\begin{equation}
\zeta\left(\br\right)=\frac{1}{V^{\frac{1}{2}}}\sum_{{\bf k}}{}^{'}\sqrt{\nu\left(\gamma,{\bf k}\right)}e^{i\left(\bar{\theta}\left({\bf k}\right)+{\bf k}\cdot\br\right)}
\end{equation}
where $\bar{\theta}\left({\bf k}\right)$ is a random phase and we
impose $\bar{\theta}(-{\bf k})=-\bar{\theta}({\bf k})$ to ensure
that $\zeta(\br)$ are real. Also, $V$ is the total volume. The prime
in the summation indicates that the ${\bf k}=0$ term is excluded
and is later added analytically. The $\gamma$-dependent long-range
Coulomb interaction in momentum space is $\nu\left(\gamma,{\bf k}\right)=4\pi e^{-\frac{|{\bf k}|^{2}\gamma^{2}}{4}}/|{\bf k}|^{2}$.
Note that the average in Eq.~(\ref{coul_decomp}) is over the random
phases $\bar{\theta}\left(\boldsymbol{{\rm k}}\right)$ which determine
the stochastic function $\zeta\left(\br\right)$.

The stochastic decompositions of the density matrix and of the Coulomb
potential are then combined to give 
\begin{align}
 & \left\langle \br\middle|X_{{\rm nl}}^{\gamma}\middle|\phi\right\rangle =\nonumber \\
 & -\left\{ \xi(\br)\langle\xi|\phi\rangle\right\} _{\bar{\eta}\theta}\simeq - \frac{1}{N_{\xi}}\sum_{j=1}^{N_{\xi}}\xi^{j}(\br)\langle\xi^{j}|\phi\rangle\label{eq:Xnl_xi}
\end{align}
where the combined exchange-operator stochastic amplitude is simply
$\xi(\br)=\eta(\br)\zeta(\br)$. The average is done now over $N_{\xi}$
random states; each sampling (labeled by $j$) of random $\xi(\br)$
amounts to a simultaneously choosing (independently) both the phases
$\theta({\bf k})$ and the random vector $\bar{\eta}({\bf \br)},$
so $\xi^{j}(\br)=\eta^{j}(\br)\zeta^{j}(\br)$. 

Eq.~(\ref{eq:Xnl_xi}) is formally exact if the number of states
$N_{\xi}\to\infty$. For any finite number $N_{\xi}$ there is a
statistical error proportional to $1/\sqrt{N_{\xi}}$, but since the
long-range exchange vector is not numerically large this error is
small even when $N_{\xi}$ is only a few hundreds. Further details,
such as the supplementary use of a deterministic HOMO/LUMO when extracting
$\gamma,$ are given in Ref.~\onlinecite{Neuhauser2015}.

Note that in this present approach the only operator which is stochastically
sampled is the long-range exchange. The density is still sampled deterministically
from the eigenstates, $n(\br)=\sum_{i}f_{i}|\phi_{i}\left(\br\right)|^{2}$,
and the DFT cost is similar to that of traditional deterministic DFT
for semilocal functionals. In practice, one can use any usual DFT
algorithm to iteratively solves $H\phi_{i}=\varepsilon_{i}\phi_{i}$
for the occupied states, with $H$ constructed from $n(\br)$ and
from $X_{{\rm nl}}^{\gamma}$ in Eq.~(\ref{eq:Xnl_xi}).

The statistical errors in this mixed approach, where only the exchange
is sampled stochastically,  
are much smaller than in our fully-stochastic DFT approach\cite{BaerNeuhauserRabani2013,Neuhauser2014a}
where the stochastic orbitals $\eta(\br)$ were also used to sample
the local density (i.e., where we use $n(\br)=\left\{ |\eta(\br)|^{2}\right\} ,$
or more generally $n(\br)=n_{0}(\br)+\left\{ |\eta(\br)|^{2}\right\} ,$
where $n_{0}(\br)$ is a fragment density). In the fully-stochastic
approach the eigenstates do not need to be determined\cite{Baer2012,BaerNeuhauserRabani2013,Neuhauser2013}
so it formally scales linearly with system size; here the scaling
goal is more modest, just to reduce the cost to that of traditional
DFT (and later TDDFT) with only local and semilocal potentials.

\subsection{Stochastic TDDFT with non-local exchange}

Following the DFT stage with stochastic exchange, we turn to the implementation
of stochastic-exchange in TDDFT. There is no need to use the same
exact methodology for the stochastic exchange in the TDDFT as in the
DFT stage. Here, we follow \cite{Rabani2015} and use a different
sampling of the stochastic density matrix at each time step. Specifically,
at each time-step we represent the density matrix as an average over
stochastic vectors, where each one is constructed from the occupied
eigenstates: 
\begin{equation}
\beta\left(\br,t\right)=\sum_{j}e^{i\alpha_{j}(t)}\sqrt{f_{j}}\phi_{j}\left(\br,t\right),\label{create_eta}
\end{equation}
where $\alpha_{i}(t)\in[0,2\pi]$ is a random phase. Thus, each $\beta$
is a stochastic vector created using a distinct set of random phases
$\left\{ \alpha_{i}\right\} $, and a different set of random phases
is taken as each time step. Obviously $\left\{ \beta(\br,t)\beta^{*}(\bp,t)\right\} =\rho(\br,\bp,t).$

Note that the $\eta$ and $\beta$ vectors have a similar meaning;
the former is used for the initial time-independent stage, the latter
for TDDFT. We use a different symbol to emphasize that the number
of such stochastic vectors is different in DFT and TDDFT. Specifically,
since each time-step is small the effect of the stochastic exchange
per time step is numerically small, so that it is sufficient to use
only a small number ($N_{\beta}$) of vectors 
in each time step. For that reason, we have not done a stochastic
resolution of the Coulomb kernel for the time-dependent exchange,
which is formally calculated now as 
\begin{align}
\left\langle \br\middle|X_{{\rm nl}}^{\gamma}(t)\middle|\phi\right\rangle  & =-\frac{1}{N_{\beta}}\sum_{l=1}^{N_{\beta}}\beta^{l}(\br,t)\nonumber \\
 & \times\int v(\br-\bp)\beta^{l*}(\bp,t)\phi(\bp)d\bp.\label{eq:Xnlt_stochastic}
\end{align}
Thus the cost of evaluating Eq.~\ref{eq:Xnlt_stochastic} is only
$N_{\beta}$-times more expensive than evaluating the Hartree term.

Since Eq.~\ref{eq:Xnlt_stochastic} is evaluated stochastically,
$H\left(t\right)$ exhibits fluctuations even when there is no perturbation.
For a linear response in $\Delta$, we therefore need to propagate
two equations, with and without the perturbation: 
\begin{align}
i\left|\dot{\phi}_{i}^{\Delta}\left(t\right)\right\rangle  & =\left[H^{\Delta}\left(t\right)+\delta v(\br,t)\right]\left|\phi_{i}^{\Delta}(t)\right\rangle \label{tprop2a}\\
i\left|\dot{\phi}_{i}^{\Delta=0}\left(t\right)\right\rangle  & =H^{\Delta=0}\left(t\right)\left|\phi_{i}^{\Delta=0}(t)\right\rangle ,\label{tprop2b}
\end{align}
where $\phi_{i}^{\Delta}\left(\br,t=0\right)=\phi_{i}^{\Delta=0}\left(\br,t=0\right)=\phi_{i}\left(\br\right)$.
$H^{\Delta}\equiv H[n^{\Delta}(t),\rho^{\Delta}(t)]$ and $H^{\Delta=0}\equiv H^{\Delta}\equiv H[n^{\Delta=0}(t),\rho^{\Delta=0}(t)]$
have the same functional dependence on the density matrix, but since
the time-dependent solutions of Eqs.~\ref{tprop2a} and \ref{tprop2b}
are different, we distinguish the Hamiltonians by superscripts. The
time evolution of $\left\{ \phi_{i}^{\Delta=0}\right\} $ stems purely
from the stochastic fluctuations in $H^{\Delta=0}\left(t\right)$,
as no external perturbing potential is applied. This fluctuation also
induces time-dependence in the charge density $n^{\Delta=0}\left(\br,t\right)$
which needs to be subtracted when calculating the induced dipole 
\begin{equation}
\mu\left(t\right)=\frac{1}{\Delta}\int(\br\cdot\hat{\boldsymbol{{\rm e}}})\left[n^{\Delta}\left(\br,t\right)-n^{\Delta=0}\left(\br,t\right)\right]\dd\br,\label{eq:dip_stoch}
\end{equation}
from which the frequency-dependent absorption follows. 

\subsection{Numerical propagation of TDDFT with stochastic exchange}

We use a split operator approach for the numerical propagation of
the TDDFT equation with stochastic exchange. As usual, the perturbation
is first applied at $t=0$ (and we again omit below the $\Delta$
superscript):

\begin{equation}
\phi_{i}(\br,t=0^{+})=e^{-i(\br\cdot\hat{\boldsymbol{{\rm e}}})\Delta}\phi_{i}(\br)
\end{equation}
and we then split the propagation of the non-local exchange and the
remainder of the Hamiltonian,

\begin{align}
|\phi_{i}(t+ & dt)\rangle=\nonumber \\
e^{-iX_{{\rm nl}}^{\gamma}\frac{dt}{2}} & e^{-i\left(h_{0}+v_{H}\left[n\left(t\right)\right]+v_{C}^{\gamma}\left[n\left({\it t}\right)\right]\right)dt}e^{-iX_{{\rm nl}}^{\gamma}\frac{dt}{2}}|\phi_{i}(t)\rangle.\label{eq:prop_1dt}
\end{align}
The short time kinetic+potential propagator (the non-$X_{{\it {\rm nl}}}^{\gamma}$
part in Eq.~(\ref{eq:prop_1dt})) is itself calculated with a usual
split operator evolution which will not be reviewed here, while $e^{-iX_{{\rm nl}}^{\gamma}\frac{dt}{2}}$ is
evaluated extremely simply as

\begin{equation}
e^{-iX_{{\rm nl}}^{\gamma}\frac{dt}{2}}|\phi\rangle\simeq N_{\phi}\left(1-iX_{{\rm nl}}^{\gamma}\frac{dt}{2}\right)|\phi\rangle\label{eq:prop_Xdt}
\end{equation}
where $N_{\phi}$ is a time-dependent normalization constant, i.e.,
$N_{\phi}^{-1}=||\left(1-iX_{{\rm nl}}^{\gamma}\frac{dt}{2}\right)|\phi\rangle||$.
Since the normalization is dependent on the initial vector, Eq.~(\ref{eq:prop_Xdt})
is slightly non-linear but this is of little practical consequence.
The primitive approach of Eq.~(\ref{eq:prop_Xdt}) is sufficient
since the time-steps used are generally small, typically $dt=0.05$~a.u.,
i.e., around 1~as.

\section{Results}

For all systems studied here, we first perform a ground-state DFT
calculation and obtain the range-separation parameter $\gamma$ by
enforcing the piecewise linearity condition for the total energy;
this ensures that the HOMO is the same as the ionization energy.

\subsection{Validation of the method using $\boldsymbol{{\rm PH_{3}}.}$}

The smallest system studied is a PH$_{3}$ molecule. Here, a deterministic
DFT calculation was performed using experimental molecular structure\cite{herzberg1966molecular}
and the valence electronic states were computed with Troullier-Martins
pseudopotentials.\cite{TroullierMartins1991} The total energy and the eigenvalues were
converged to 5meV with a real space grid of $64\times64\times64$
points and a $0.4\,a_{0}$ grid spacing. Note that small molecular
systems require, in general, a large range separation parameter and
converge slower with the grid size and spacing compared to large systems.
Through the tuning procedure,\cite{Stein2010,Kronik2012,Neuhauser2015}
we found that $\gamma=0.37\,a_{0}^{-1}$; the resulting ionization potential
(i.e., the negative of the HOMO energy) is $10.4$~eV, in excellent agreement with
experiment ($10.6$~eV - Ref.~\onlinecite{cowley1982lewis}). 

The LUMO (obtained with the same range separation parameter) is barely
bound, by slightly less than 0.1eV, but experimentally PH$_{3}$ does
not form a stable anion so the LUMO energy should be non-negative.

\begin{figure}
\centering \includegraphics[width=0.5\textwidth]{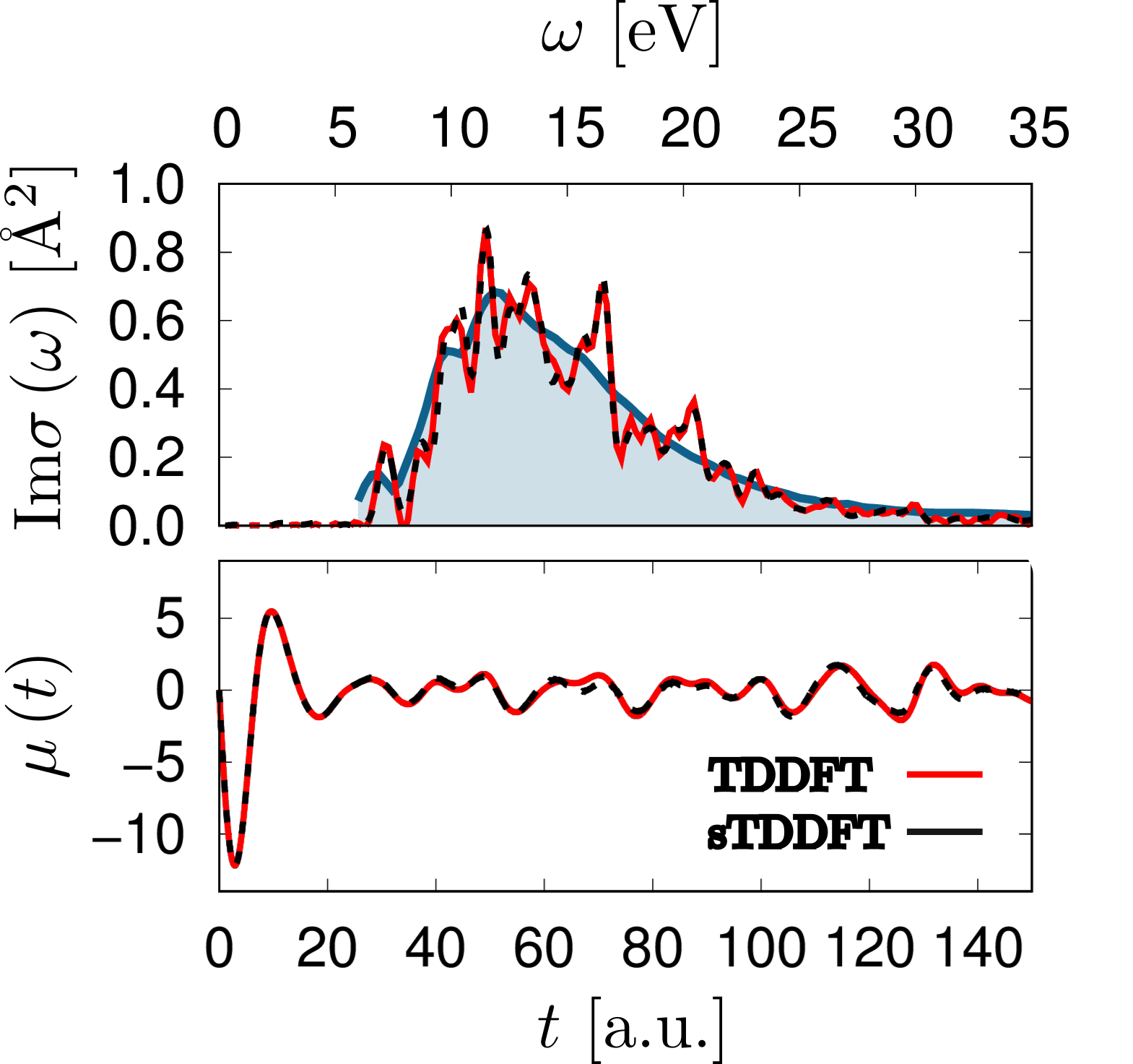}
\caption{\textbf{Top:} Optical absorption of a PH$_{3}$ molecule from stochastic
and deterministic TDDFT simulations (full and dashed lines, respectively).
The experimental spectrum \cite{zarate1990absolute} is shown by a
blue shaded area. \textbf{Bottom:} An initial segment of the time
propagation showing the evolution of the induced dipole $\mu\left(t\right)$. }
\label{PH3-comparison} 
\end{figure}

Using the optimally tuned BNL functional, the optical cross-section
$\sigma\left(\omega\right)$ was obtained (see Eq. (\ref{cross2}))
by deterministic and stochastic real time propagations, and the results
are shown in Fig.~\ref{PH3-comparison}. The TDDFT equations were
propagated for a total time of 24~fs which provides a spectral resolution
of $\sim$170~meV. The computed absorption cross-section has a first
peak at $E_{1}=7.1$~eV, in excellent agreement with the first experimental
peak at $7.0$~eV.

The exciton binding energy is defined as 
\begin{equation}
E_{b}=E_{g}-E_{1},\label{exciton_binding_energy}
\end{equation}
where $E_{g}$ is the fundamental band gap taken as the difference
between HOMO and LUMO energies, i.e., $E_{g}=10.3$ eV. The predicted
PH$_{3}$ binding energy is thus $E_{b}=3.2$~eV, close to the experimental
value of $3.4$ eV.\footnote{For PH$_{3}$, we assume that the fundamental band gap coincides with
the ionization potential as the molecule does not bind an extra electron}

The overall absorption maximum is at $11.5$~eV, in good agreement
with experiment ($12.0$~eV), though the latter exhibits large peak
widths. At higher frequencies, the TDDFT spectrum has multiple local
maxima (e.g., at $16.5$ and $20.4$~eV) that in the experiment only
appear as shoulders. This is because experimental measurements cannot
be resolved at energies higher than the ionization threshold ($10.4$~eV).

The stochastic decomposition of the time-dependent exchange (sTDDFT
- Eq.~\ref{eq:Xnlt_stochastic}) reproduces the deterministic results
already for $N_{\beta}=2$ (note that this is half the number used
in deterministic exchange, which involves four valence states). We
also checked that the spectrum does not change when the perturbation
is varied in the range $\Delta=10^{-4}-10^{-3}$~a.u., as expected
in linear response; this was also checked for the phosphorene sheets,
discussed next.

\subsection{Phosphorene Sheets: DFT}

The major advantage of stochastic approaches is their applicability
to large systems. We demonstrate this feature now on a set of 2D phosphorene
sheets of increasing sizes, derived from a black phosphorus crystal
structure.\cite{Cartz1979} The sheets were passivated with H atoms
on the rims; each P atom that would have been bound to two H atoms
was removed, resulting in a compact sheet geometry. Note that the
two in-plane directions in the phosphorene sheet are traditionally
labeled as armchair and zig-zag. With a kinetic energy cutoff of 26 $E_h$ and a real space grid with 208$\times$136$\times$40 points
and a 0.6~$a_{0}$ spacing, the Kohn-Sham eigenvalues were converged
to 10 meV. 

The smallest sheet is $0.6\times0.8$~nm and has 112 valence electrons.
For its ground state DFT calculations we employed both the deterministic
and stochastic formulation of the exchange operator $X_{{\rm nl}}^{\gamma}$. The optimally tuned range-separation parameter for this sheet is
$\gamma=0.10\,a_{0}^{-1}$, and the stochastic eigenvalues converge
slowly with the number of stochastic states, so $N_{\xi}\sim1600$
is required to yield a statistical error of $<0.05$~eV.

In addition to the small sheet, we considered two larger sheets, $1.3\times2.1$~nm
and $3.1\times4.3$~nm (labeled ``medium'' and ``big''), with
478 and 1958 valence electrons respectively. For these larger sheets,
the exchange operator was calculated purely stochastically as the
deterministic calculation would have been very expensive. The range
separation parameter gradually decreases with system size as in other
1D and 3D systems,\cite{Stein2010,Korzdorfer2011,Vlcek2016} so $\gamma=0.09a_{0}^{-1}$
in the medium-size sheet and $0.05a_{0}^{-1}$ in the largest one.
As the range separation parameter decreases with system size, the
long-range exchange operator $X_{{\rm nl}}^{\gamma}$ is numerically
smaller and its stochastic representation has therefore a small absolute
statistical error. Hence, the largest system requires a smaller value
of $N_{\xi}$ in the ground state calculations. Namely, the eigenvalues
are converged to $<50$~meV with $N_{\xi}=1600$ stochastic states
for the medium-size sheet and only $N_{\xi}=400$ for the largest
one.

The stochastic-exchange DFT yields fundamental gaps $E_{g}$ that
decrease with system size: $E_{g}=3.9$, $3.1$ and $1.7$~eV for
the three sheets respectively. {The large-sheet result is very similar to
the HSE hybrid functional prediction of $1.5$~eV.\cite{Qiao2014} Comparing
to with previous periodic $G_{0}W_{0}$ calculations (with a PBE starting
point), we find that the largest sheet is in rough but not perfect
agreement with the 2D periodic $G_0W_0$ fundamental
gap of 2.08 eV.\cite{li2017direct,qiu2017environmental} The difference
from the $GW$ result could be due to the scalar (i.e., non-directional)
nature of the range separated parameter, which ignores the difference
between the effective interactions in the in- and out-of plane directions
\cite{vlcekneuhauser2b}, and perhaps also due to the approximate
nature of the $G_{0}W_{0}$ itself.

\begin{figure}
\centering \includegraphics[width=0.5\textwidth]{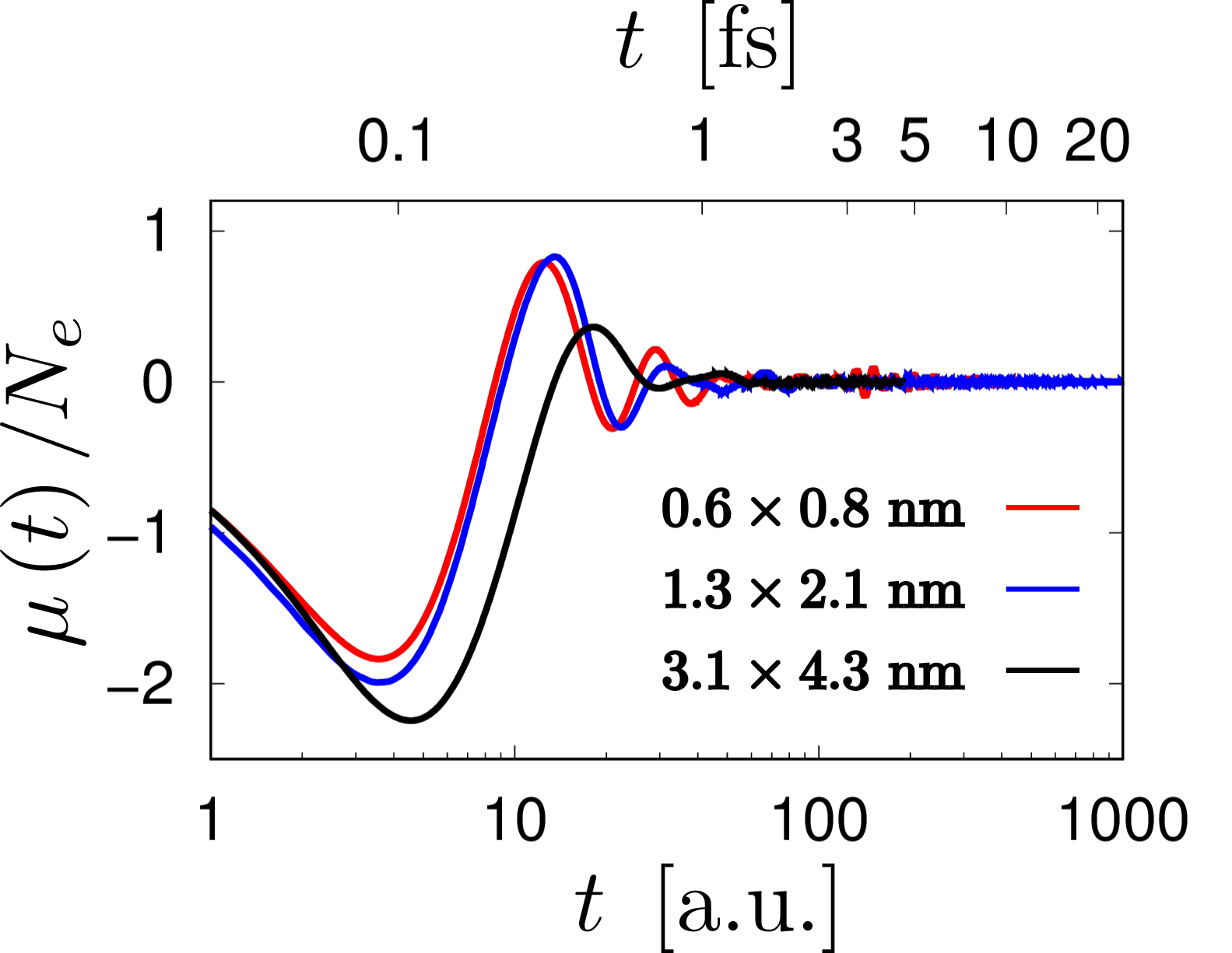}
\caption{Induced dipole per electron, $\mu\left(t\right)/N_{e}$, along the
armchair direction of phosphorene, plotted for three phosphorene sheets
with different lateral dimensions. The largest system (black) was
propagated only to 210 a.u., after which the stochastic fluctuations
dominate the signal. The periods of the induced dipole oscillations
grow with system size. A logarithmic time axis is used; the oscilations
are mostly non-stochastic and due to the logarithmic axis.}
\label{tprop_sheets} 
\end{figure}

\subsection{Phosphorene sheets: TDDFT results}

We next discuss the stability of TDDFT simulations for the sheets
and the resulting optical spectra. Since the wavefunction is only
incremented gradually, by $dt/2=0.025$~a.u, the statistical fluctuation
introduced by each stochastic decomposition of the exchange operator
is significantly smaller than for the ground state calculation. Hence,
as mentioned, a small $N_{\beta}$ is sufficient for the time-dependent
calculation, so the short-time results are fairly accurate already
for $N_{\beta}=1-4$ . However, $N_{\beta}$ influences the total
time of the simulations since due to statistical
fluctuations the propagation eventually becomes unstable. We verified
that this instability is not influenced by the time step, grid size and
kinetic-energy cut-off.

We noticed the instability phenomena already in our original stochastic
TDDFT approach\cite{Gao2015} where, unlike here, we propagated only
a few stochastic combination of eigenstates (i.e., several $\beta(\br,t))$,
and constructed from them the density as $n(\br,t)\simeq N_{\beta}^{-1}\sum_{\beta}|\beta(\br,t)|^{2}$).
That approach is extremely efficient for short time simulations (where
the plasmon response of systems with thousands of electrons is accurately
modeled by circa ten propagated states), but is limited to short times
since the propagation eventually becomes unstable. Here, since all
occupied eigenstates are propagated, and the density is constructed
from all of them, the propagation is fairly stable for longer times.

Specifically, for the small and medium sheets, with 112 and 478 valence
electrons, the propagation was carried up to $1000$~a.u ($\sim24$~fs)
without stability issues, using $N_{\beta}=$2 . This is in line with
our previous simulations (Ref.~\onlinecite{Rabani2015}) which used
a damped exchange (reduced by 80\%) in 3D and were stable with $N_{\beta}=1$.
However, for the large sheet (with 1958 electrons) the time evolution
became numerically unstable after $260$~a.u. ($\sim7$ fs) even
with $N_{\beta}$=6. The instability is for two reasons; first, the
rapid oscillations of the density in the direction perpendicular to
the phosphorene sheet. Unlike 3D systems, the response here is highly
anisotropic and this appears to enhance the stochastic noise in the
time propagation. Hence, for large 2D sheets the value of $N_{\beta}$
needs to be increased. Further, the form of Eq. (\ref{eq:Xnlt_stochastic})
is oscillatory even for $\Delta=0$; in future publications we would
use a less oscillatory form analogous to that in Ref.~\onlinecite{Rabani2015}
and, in addition, would use a fully separable form of the TDDFT calculations,
analogous to Eq. (\ref{eq:Xnl_xi}) for DFT with exchange.

\begin{figure*}
\centering \includegraphics[width=0.85\textwidth]{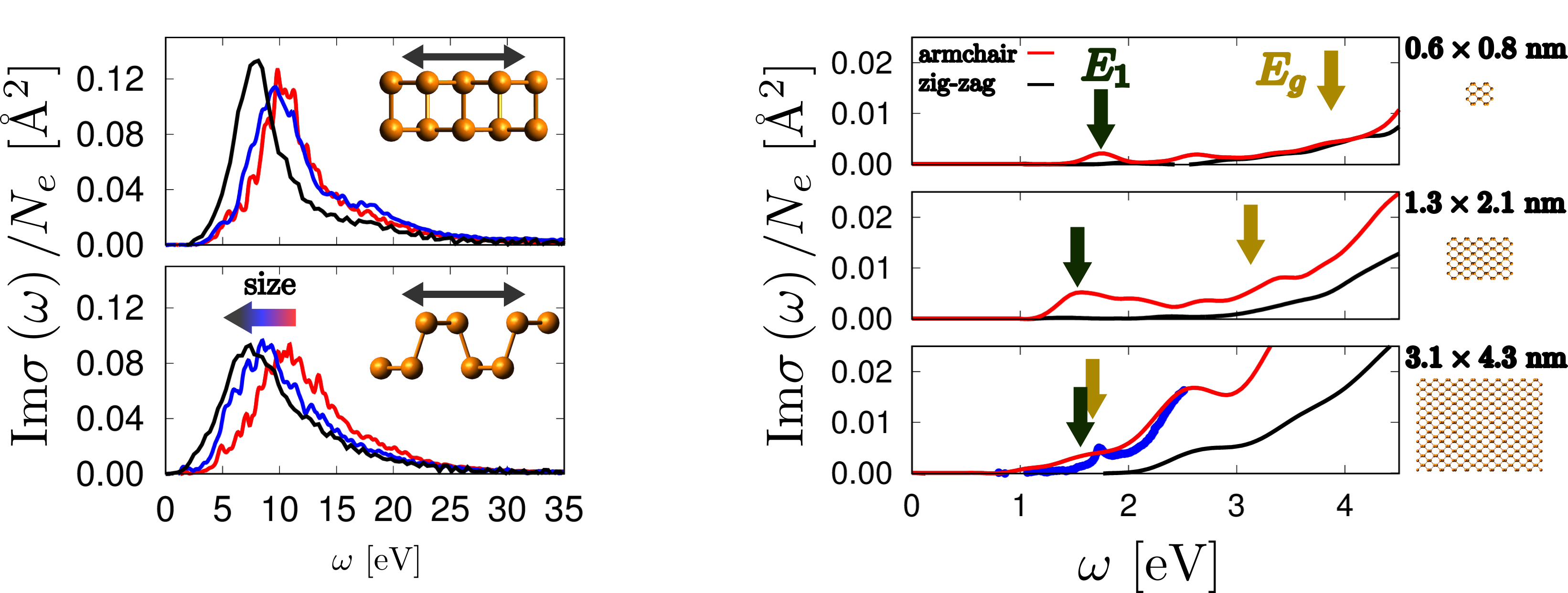}
\caption{The left panel shows the theoretical optical spectra of phosphorene
sheets of increasing sizes: $0.6\times0.8$~nm (112 valence electrons),
$1.3\times2.1$~nm (478 valence electrons) and $3.1\times4.3$~nm
(1958 valence electrons) marked by red, blue and black lines respectively
for the zig-zag (top) and armchair (bottom) directions. The right
panel shows details of the low energy portion of the spectra; the
yellow and dark-green vertical lines indicate the positions of the
fundamental band gap ($E_{g}$) and the first excitonic peak ($E_{1}$).
Experimental data for a bulk 2D monolayer phosphorene (taken from
Ref.~\cite{li2017direct}) are shown in the bottom graph by blue
points. Features that lie between $E_{1}$ and $E_{g}$ correspond
to multiple excitonic states. }
\label{spectra_big} 
\end{figure*}

Next, we turn to discuss the individual results. The time evolution
of the induced dipole for the three phosphorene sheets is shown in
Fig.~\ref{tprop_sheets}. The absorption cross sections per-electron
are shown in Fig.~\ref{spectra_big} for the zig-zag and armchair
directions. The spectra are strongly anisotropic, in agreement with
experimental data and $GW$/BSE calculations on 2D periodic sheets.\cite{liu2014,Xia2014,Wang2015a,li2017direct}
With rising system size there are diminished quantum confinement effects
so the fundamental band gap decreases and the absorption maximum therefore
gradually shifts to lower energies. As the number of valence electrons
increases the spectra also smoothens.

On the right panel of Fig.~\ref{spectra_big} we zoom on the absorption
spectrum below the ionization potential (which only slowly decreases
with system size, from 5.9 to 5.4 eV). The absorption cross section
decreases rapidly at lower frequencies, but several local maxima are
evident. Some of these local maxima are below the band gap energy
($E_{g}$) and therefore correspond to bound electron-hole pairs.
These excitonic peaks appear only for the armchair direction (due
to the strong anisotropy of the optical response); this feature was
seen in previous calculations for periodic phosphorene and was also
seen experimentally.\cite{liu2014,Tran2014,li2017direct}

Excitonic peaks are usually sharp and have a high intensity, indicating
long-lived quasiparticle states. The maxima in Fig.~\ref{spectra_big}
are however broadened due to the finite simulation time (24 fs for
the two small systems and 7 fs for the largest one). For the small
sheet, the excitonic peaks are well-separated but have relatively
low intensity. The position of the first absorption peak maximum ($E_{1}$)
changes with increasing system size from $1.8$ to $1.6$~eV. The
latter is in good agreement with the experimental value and $GW$/BSE
estimates, $1.7$ and $1.6$~eV, respectively \cite{li2017direct,qiu2017environmental}
for a bulk 2D system.

The exciton binding energy (Eq.~\ref{exciton_binding_energy}) decreases
rapidly with system size from $2.1$ to $0.1$~eV. The strongest
excitonic response (the largest amplitude of the $E_{1}$ peak) is
found in the medium sized system, which also has a high exciton binding
energy $E_{b}=1.65$~eV. As mentioned, however, the fundamental gap
$E_{g}$ in the stochastic-exchange DFT is underestimated relative
to $G_{0}W_{0}$ calculations. Therefore, the exciton binding energy
for the largest sheet (0.1~eV) is much lower than predicted by $GW$/BSE
calculations which give $E_{b}=0.48$~eV.\cite{qiu2017environmental}

Interestingly, when the phosphorene is encapsulated in dielectric
media, the $GW$/BSE binding energy becomes small, $0.14$~eV \cite{qiu2017environmental},
comparable to our TDDFT estimates of pristine (non-encapsulated) phosphorenes.
The encapsulation causes strong screening above and below the 2D system
(i.e., in the out-of-plane direction) leading to a big change in the
$G_{0}W_{0}$ gap $E_{g}$ (from $2.08$ to $1.62$~eV, similar to
our pristine DFT gap), while the position of the first excitonic peak,
$E_{1}$ remains practically unaffected.\cite{qiu2017environmental}
The difference between our results and experiment and $GW$/BSE points
to a problem in describing 2D materials with range-separated potentials,
\cite{vlcekneuhauser2b} which should in principle account for the
anisotropy of the electron-electron interaction between the in- and
out-of-plane directions. Until such a non-isotropic interaction is implemented in DFT and TDDFT,
stochastic TDDFT with exchange can only be trusted as far as the exciton
frequency, but calculations of the exciton binding would require a more accurate method than GKS-DFT for the quasiparticle gap.
\section{Conclusions}

In summary, we developed an efficient real-time TDDFT approach with
stochastic long-range non-local exchange. The stochastic treatment
decomposes the density matrix in TDDFT to an average over a product
of random vectors $\beta$ in the space spanned by the occupied orbitals.
It significantly reduces the computational cost as only a few stochastic
states are needed at each time step. Further, the number of stochastic
states varies only a little with the system size. Calculations for
very large systems thus become feasible.

The resulting TDDFT with long-range non-local exchange includes the
attractive electron-hole interaction that gives rise to exciton formation.
Indeed, our TDDFT yields optical spectra that are in excellent agreement
with experiment. For small systems, where deterministic calculations
are affordable, the stochastic and deterministic results agree.

We demonstrated that our method is applicable for extremely big systems
using a set of phosphorene sheets containing up to $\sim$2000
valence electrons. The largest system was compared to experiments
and previous calculations on infinite phosphorene sheets; the analysis
confirms that the range-separated hybrid functional successfully predicts
optical spectra even with strong excitonic signatures.

\section{Acknowledgments}

D.N. acknowledges support by NSF grant CHE-1763176. RB acknoledges the US-Israel Binational Fund
grant no. BSF2015687. The authors would like to acknowledge helpful discussions with Eran Rabani.
The calculations
were performed as part of the XSEDE\cite{towns2014xsede} computational
project TG-CHE180051. 
\section*{Appendix I: Photoabsorption cross section}

\setcounter{equation}{0} \global\long\def\theequation{A.\arabic{equation}}

Here we overview for completeness the well-known expression of the
photoabsorption cross-section as a Fourier transform of a real-time
dipole correlation function.

The absorption cross section, $\sigma(\omega)$, is given in linear
response as \cite{OnidaReiningRubio2002}: 
\begin{align}
 & \sigma\left(\omega\right)=\nonumber \\
 & \frac{4\pi}{c}\omega\iint\delta\tilde{v}\left(\br,\omega\right)\tilde{\chi}\left(\br,\bp,\omega\right)\delta\tilde{v}\left(\bp,\omega\right)\;\dd\br\;\dd\bp,
\end{align}
where tilde is used occasionally to denote quantities in frequency
domain, $\delta\tilde{v}\left(\br,\omega\right)$ is dynamical external
potential and $\chi$ is the electronic reducible polarizability,
which is given in the time domain as: 
\begin{equation}
\chi\left(\br,\bp,t-t^{\prime}\right)=\frac{\delta n\left(\br,t\right)}{\delta v\left(\bp,t^{\prime}\right)},\label{polarizability}
\end{equation}
where $\delta n\left(\br,t\right)$ is the induced charge density
at a point $\br$ and time $t$. The response function is causal,
i.e., $t>t^{\prime}$. For absorption of polarized light along a unit
vector $\hat{{\rm \boldsymbol{e}}}$ we apply $\delta v$ as a dipole
potential. The cross section is then \cite{OnidaReiningRubio2002}
\begin{equation}
\sigma\left(\omega\right)=\frac{4\pi}{c}\omega\iint(\hat{{\rm \boldsymbol{e}}}\cdot\br)\tilde{\chi}\left(\br,\bp,\omega\right)\cdot(\hat{{\rm \boldsymbol{e}}}\cdot\boldsymbol{\bp})\;\dd\br\;\dd\bp.\label{cross1}
\end{equation}

Here, $\sigma$ is calculated from real-time linear-response. Specifically,
the first stage is to apply a dipole external potential perturbation
\begin{equation}
\delta\nu(\bp,t)=(\hat{{\rm \boldsymbol{e}}}\cdot\bp)\delta\left(t\right)\Delta,\label{perturbation}
\end{equation}
where $\Delta$ is the perturbation strength, and an instantaneous
perturbation is applied at $t=0$ allowing to probe the response at
all frequencies. The perturbation potential is applied to all occupied
eigenstates, which are then propagated in time. The resulting oscillations
of the induced charge density (Eq.~\ref{polarizability}) are then
used to find the dipole auto-correlation, 
\begin{equation}
\mu(t)=\frac{1}{\Delta}\int(\hat{{\rm \boldsymbol{e}}}\cdot\br)\cdot\delta n\left(\br,t\right)\;\dd\br.\label{dipdip}
\end{equation}
The absorption cross section is finally a Fourier transform of the
dipole auto correlation: 
\begin{equation}
\sigma\left(\omega\right)=\frac{4\pi\omega}{c}\int_{0}^{\infty}\mu(t)e^{i\omega t}\;{\dd}t.\label{cross2}
\end{equation}

\bibliographystyle{aipnum4-1}
\bibliography{library-stochX}

\end{document}